\documentclass[showpacs,twocolumn,pre]{revtex4}
\usepackage{graphicx, amsmath, amsthm, amssymb}
\usepackage{xcolor}

\DeclareUnicodeCharacter{2212}{-}

\begin{document}
\title{Mechanism of laser induced void array formation in Polydimethylsiloxane (PDMS)}
\author{N. Naseri$^1$, A. M. Alshehri$^3$, L. Ramunno$^2$,R. Bhardwaj$^2$}
\affiliation{$^1$Department of Physics, Adelphi University, One South Avenue,  
Garden City, NY 11530-0701, U.S.A}
\affiliation{$^2$ Department of Physics, University of Ottawa and Nexus for Quantum Technologies Institute, 25 Templeton St., Ottawa, Ontario K1N 6N5, Canada}
\affiliation{$^3$ Department of Physics, King Khalid University,  P.O. Box 9004, Abha, Saudi Arabia}
 
\begin{abstract}
This study investigates the formation of multi-voids in polydimethylsiloxane (PDMS) using a multi-pulse irradiation method and explored the impact of laser energy, number of pulses per micron (writing speed), and laser spot size on the process. Our experimental results shows that multi-void formation occurred due to multi-pulse irradiation in the bulk of PDMS. Additionally, increasing laser energy led to an increase in the number of voids, while the number of voids did not change with an increase in the number of pulses per micron for  fixed laser parameters. However, the size of the voids increased with the number of pulses per micron, and tighter focusing conditions resulted in smaller voids with a shorter distance between them.
Furthermore, Finite-Difference-Time-Domain (FDTD) simulations reproduced the generation of void arrays in PDMS using a similar multi-laser pulse approach. We modeled the voids as concentric spheres with densified shells and implemented the pre-recorded void(s) in the medium. By studying the laser interaction with the implemented void(s), we observed electric field enhancement in front of the void(s), resulting in the generation of subsequent void. Our study shows that the generation of void arrays in PDMS follows a linear mechanism. This study provides valuable insight into the mechanism behind the formation of void arrays in PDMS. The simulation results agree with the experimental results to further validate the model and gain a better understanding of the physical processes involved in the generation of void arrays in PDMS.
\end{abstract}

\maketitle
\section{Introduction}
Femtosecond laser modification and damage to transparent materials has been an active research area for decades due to numerous  promising applications in different areas of photonics \cite{Gattas,Zhang2016,Beresna2017,Drev}. Interaction of femtosecond laser pulses with transparent materials and polymers can lead to various processes depending on laser intensity and laser spot size. 
Among these modifications, void formation induced by femtosecond laser pulses has received significant attention due to its potential applications in devices such as optical memories, waveguides, gratings, couplers, and chemical and biological membranes\cite{Graf,Beresna2017,Bell,Davis,Shimotsuma,Zhang,Zhang2016,Beresna,Drev}. 
The fabrication of voids in silica glass \cite{Glezer1996,Glezer1997,Watanabe1999,Watanabe2000,Toratani2005,Dai2016} and polymers \cite{Yamasaki2000,Day2002} has been widely reported in numerous studies since the novel work of Glezer et al. \cite{Glezer1996}. The shape of the damage structure in transparent materials is determined by the laser focusing condition. Numerical and experimental studies conducted over the past two decades on single-pulse filamentation in transparent materials have demonstrated that the femtosecond laser interaction with the material is heavily influenced by the laser focusing conditions. Tight focusing produces voids \cite{Glezer1997,Schaffer2001}, while loose focusing creates long channels of modified refractive index (See. Naseri, et. al \cite{Naseri} and references therein).
 Loose laser focusing refers to laser spot size few times larger than the laser wavelength. In this regime, the dominant process is the nonlinear Kerr effect, which leads to the formation of filaments as the laser pulse interacts with the material. In contrast, tight focusing is characterized by laser spot sizes comparable to or smaller than the laser wavelength, with optical focusing as the dominant mechanism. This results in the formation of a void structure near the geometrical focus position \cite{Naseri}. The number of laser pulses that interact with the material also plays a significant role in determining the outcome of the laser-material interaction.\\
 A void structure is typically characterized by a central volume of less dense material that is surrounded by a higher density material \cite{Glezer1996, Glezer1997}. Despite extensive research dedicated to elucidating the mechanisms behind multi-void formation in materials, the underlying aspects and mechanism of self-void array formation in dielectrics and polymers remain poorly understood.\\
Previous studies have suggested that self-focusing \cite{Kanehira} and spherical aberration \cite{Song, Wang} are mechanisms responsible for laser-induced multi-void formation. However, the laser re-focusing period \cite{Wu} is much longer than that of the void arrays, and it only occurs under loose laser focusing conditions, which is not the case for multi-void formation. As a result, nonlinear self-focusing is unlikely to be the mechanism responsible for the formation of void arrays. Gaining a better understanding of the mechanisms underlying void array formation in materials could enable higher precision micromachining and more precise control over induced material changes. Polydimethylsiloxane (PDMS) is an optically transparent and a biocompatible material that can be cast into different shapes and sizes using molds with relative ease.
\\
In this paper, we show experimentally that multi-void formation is primarily a result  multi-pulse interaction with bulk PDMS under tight focusing conditions. We investigate the effects of laser pulse energy, number of laser pulses, and laser spot size on the size and number of voids in an array generated in PDMS. The size of the voids decrease with the progress of the pulses inside the medium. 
Numerically, we show that a single tightly focused laser pulse can generate a void structure in the bulk of PDMS.  To study the multi-pulse effect, we modeled the first damage structure (void) as pre-recorded into the medium and study the interaction of a subsequent laser pulse with pre-existing void. We show the generation of a second void as a result of field enhancement in front of the first void. We then model the two generated voids as pre-recorded in the medium and show the generation of the third void. Our method shows that this can continue, and successive voids can be generated as a result of multi-shot laser interaction in PDMS.
 Our model eliminates the effects of spherical aberrations and  confirms that  multi-void formation is a multi-pulse effect, and is mostly a linear mechanism.\\
\section{Experimental Setup and Results}
Femtosecond pulses from a Ti:Sapphire regenerative amplifier, operating at a repetition rate of 1 kHz, 800 nm, and 40 fs with maximum output energy of 2.5 $mJ$, were focused 300 $\mu m$ below the surface of PDMS by 0.50 and 0.65 numerical aperture (NA) aspheric microscope objectives. PDMS sample was placed on three-axis translation stages with a resolution of 50 nm along the lateral dimensions (X,Y) and 100 nm along the axial direction (Z). PDMS is an optically transparent and inert material, widely used in photonic applications owing to its excellent chemical stability and moldability, and being inexpensive.
The position of the laser focus relative to the surface of bulk PDMS sample was precisely determined by imaging the back-reflected light with a CCD camera at low pulse energies, below the ablation threshold. By translating the PDMS at different speeds the number of laser pulses incident per micron of the sample was varied.  A combination of a half-wave plate and polarizer were used to vary the pulse energy. The incident pulse energies were measured after the microscope objective and monitored by a calibrated fast photodiode operating in linear regime. The pulse duration was measured to be 65 $fs$ at the back aperture of the objective after passing through all the optics by means of a single-shot autocorrelations.  
\begin{figure}[t!]
\centering
\includegraphics[width=\columnwidth,trim=0.2cm 0cm 0cm 0cm, clip]{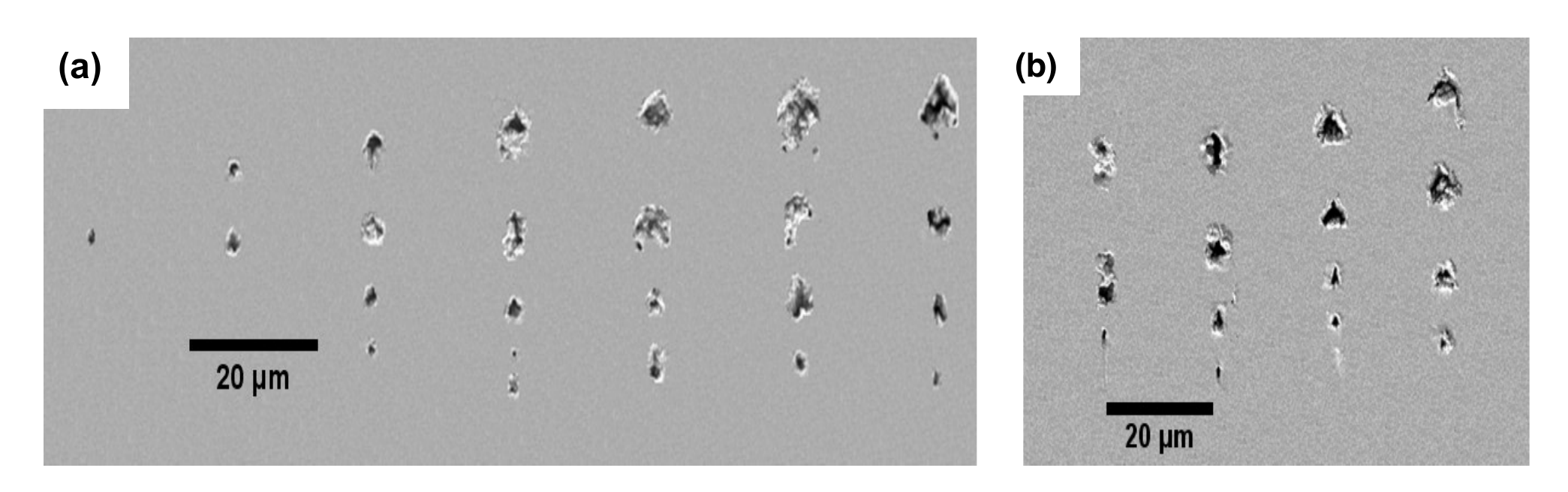} 
\caption{ SEM images of multi void structures formed with 0.5 NA aspheric objective. (a) Foci structure at different pulse energies of 50, 80, 130, 190, 260, 330, and 420 nJ, with a writing speed of 0.1mm/s.
(b) Foci structure at a fixed energy of 100 nJ with different writing speeds of 0.5, 0.08,  0.05 and  0.01 mm/s corresponding to 2, 12 , 20 , and 100 laser pulses/micron, respectively.}
 \label{fig1e}
\end{figure}
PDMS samples were prepared by mixing Dow Corning Sylgard 184 silicone elastomer base with curing agent in 10:1 ratio. The mixture was stirred for 10 minutes and degassed by placing it in a vacuum desiccator for 30 minutes. The mixture was then poured between four glasses slides that were attached to a silicon wafer that was cleaned by methanol. Another glass slide was used to gently spread the PDMS mixture avoiding any air bubbles. A weight of 500 g was positioned on the top of the glass slide to ensure flat top and the bottom surfaces. The mixture was cured in the oven at 80◦C for two hours. Large sheets with dimensions of $ 25 mm \times 75 mm \times 1mm$ were prepared that were subsequently cut into smaller samples of $10 mm \times 10 mm \times 1mm$. After laser modification of PDMS the samples were cleaved and gold coated for characterization with scanning electron microscope (SEM).\\
To investigate laser induced modification of PDMS, lines were fabricated with different pulse energies, speeds and laser spot sizes. The samples were then cut into two pieces perpendicular to the scan direction of the laser beam. The inside surfaces were characterized by Scanning Electron Microscope(SEM) as shown in figs. {\ref{fig1e},\ref{fig2e}. 

We first investigate the effect of laser pulse energy. Figure \ref{fig1e}-a shows SEM images of the single and multi-void formation for varying laser pulse energy when femtosecond pulses were focused by a 0.5 NA  objective (laser spot size of $0.5~\mu m$) and a fixed writing speed of 0.1mm/s corresponding to an irradiation by 10 laser pulses per micron. When the pulse energy is close to the threshold energy (threshold energy for NA=0.5 is $80~nJ$), a single void was formed. Increasing the pulse energy led to formation of quasi periodic-self-formed voids. In addition, void size (ones closer to the surface where light was incident) and their spacing increased with pulse energy.
Irradiation of PDMS with a single femtosecond pulse of different energies produced a single void. Figures \ref{fig1e},\ref{fig2e} reveal that the multi-void structure is a consequence of irradiating PDMS with multiple laser pulses irradiation. For a fixed pulse energy, increasing the number of laser pulses (fig. \ref{fig1e}-b) marginally increased the number of voids. However, the size of the voids increases with the number of pulses per micron. The SEM images also reveals that the position of the first void is back shifted with the increase of peak power of the laser pulse and number of pulses per micron as shown in fig.\ref{fig1e}-a,b. 
Figure \ref{fig2e} shows that besides the number of laser pulses and pulse energy, the numerical aperture of the microscope objective (laser spot size) also affects the number of voids per a single array. Focusing femtosecond pulses with a 0.68 NA objective (laser spot size of $2.8~\mu m$) resulted in 7-10 voids per single array in contrast to a maximum of four voids formed with 0.5 NA objective. In addition, the back shifting of the first void is not as dramatic as with 0.5NA objective. Also, the void sizes and the distance between them are smaller. Unlike the results of 0.5 NA objective, the number of laser pulses appear to have more significant influence on multi-void formation.
 \begin{figure}
\centering
\includegraphics[width=\columnwidth,trim=0.2cm 0cm 0cm 0cm, clip]{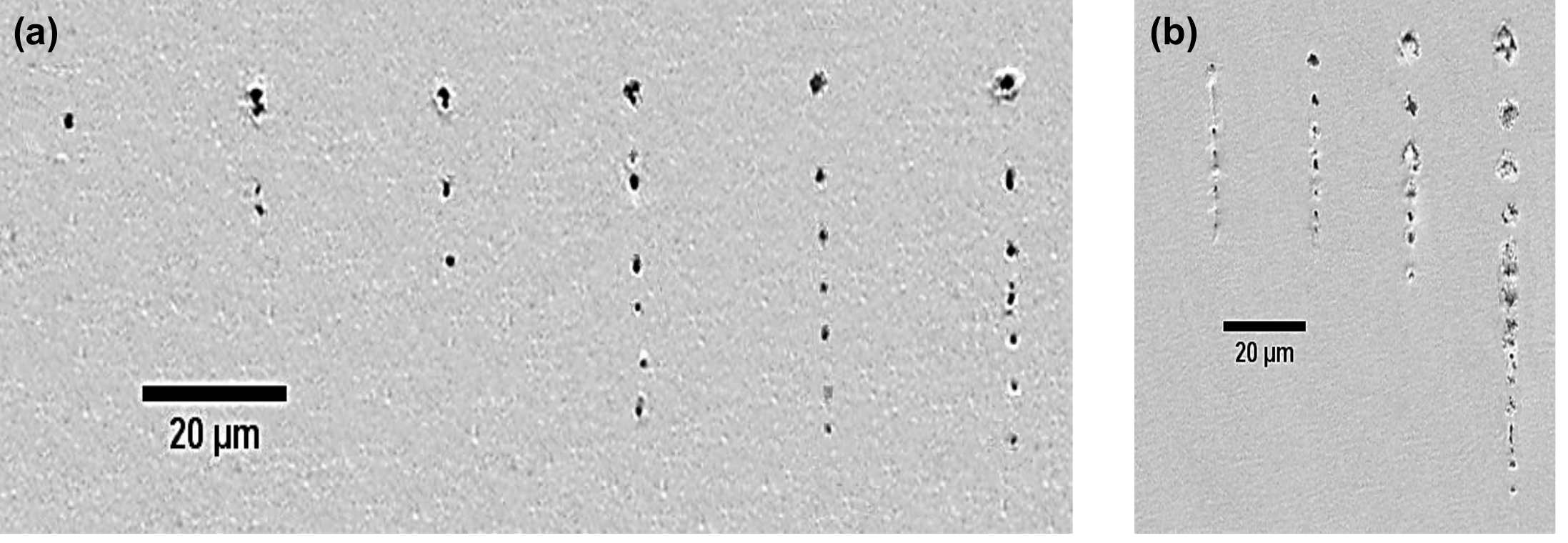} 
\caption{ SEM images of multi void structures formed with 0.68 NA aspheric objective. (a) Voids at different pulse energies of 45,70,80,85,90 and 100 nJ, with a writing speed of 0.1mm/s.  (b) Voids at a fixed energy of 85 nJ and different writing speeds (1.0, 0.4,  0.05  0.01) mm/s corresponding to 1, 12 , 20 and 100 laser pulses/micron, respectively. }
\label{fig2e}
\end{figure}

Figure \ref{fig3e} compares the size of the voids of different orders with increasing the laser pulse energy. The first order void is the one closest to the surface on which light is incident. For certain laser peak power, the size of the voids decreases with their order; first generated void is the largest, while fourth void is the smallest. Figure \ref{fig3e} also shows that for each void order, increasing the laser peak power results in larger void sizes. For example, the first generated void is the largest for the highest laser peak power, and this behavior holds for the second, third and fourth generated voids.\\  

\begin{figure}
\includegraphics[ width=0.4\textwidth]{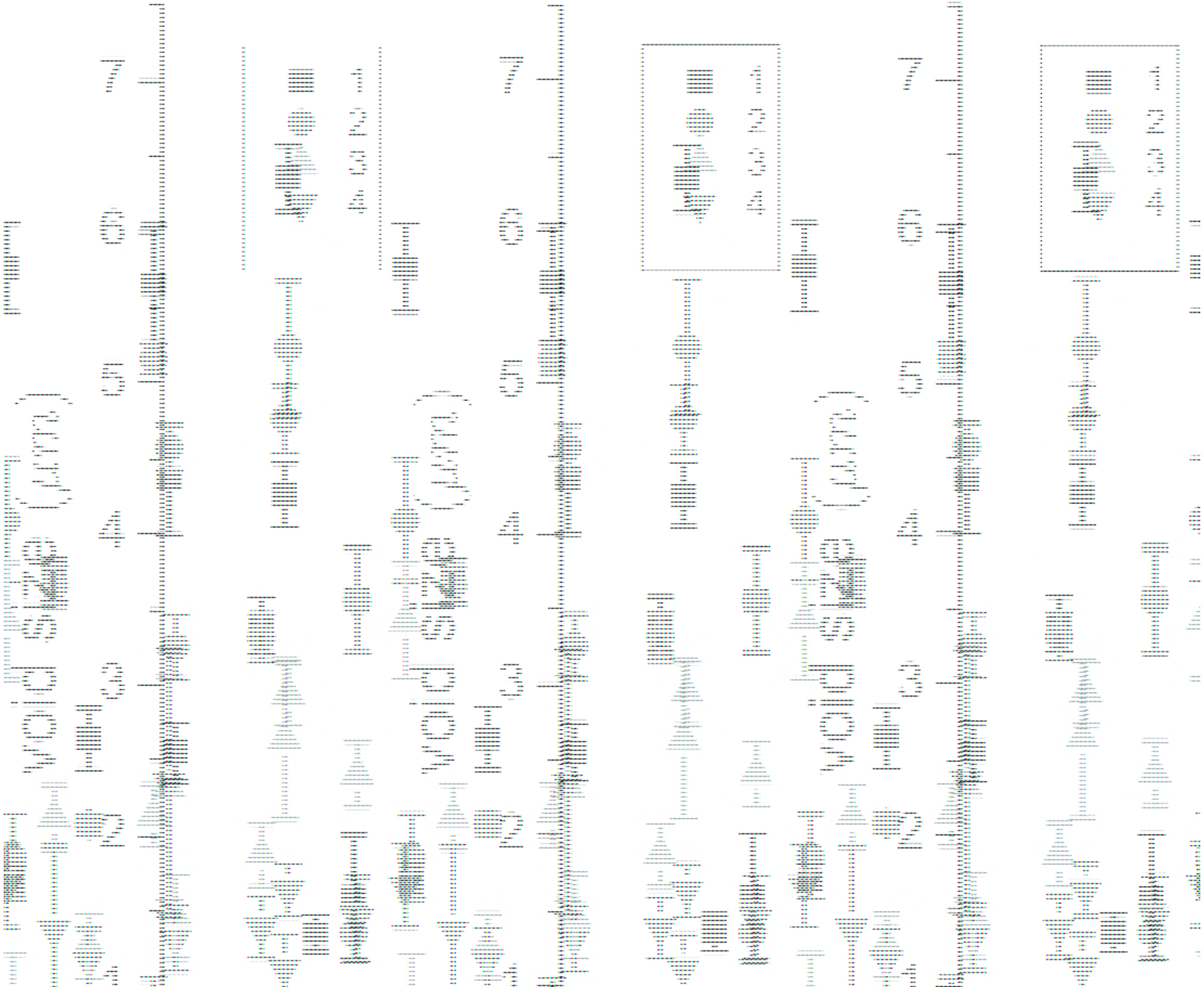} 
\caption{Size of different orders of voids as a function of peak laser power with a writing speed 
of 0.1 mm/s corresponding to 10 pulses /micron (NA=0.50). First order void is the closest to the surface on which the light is incident. Numbers 1, 2, 3, and 4 in the legend represents first, second, third and fourth  voids from the surface}.
\label{fig3e}
\end{figure}
\section{Numerical Method and Results}

To understand the mechanism of multi-void formation and our experimental results, we performed three-dimensional high resolution finite-difference-time-domain (FDTD) simulations. To date, such simulations relied on nonlinear pulse propagation with plasma generation and dynamics, using approximations such as slowly varying envelopes or evolving waves, to obtain an evolution equation for the pulse \cite{Cou}. This method has proven to be effective when the right approximations are made for the specific problem, such as in cases where the focusing is not too tight, the medium is relatively uniform, and the plasma created is not too dense \cite{Dub}. However, since the spot size of the laser and the void size in materials are usually a fraction of the laser wavelength, a more rigorous computational approach that does not make any assumptions about light propagation is required \cite{Popov, Bulga, Rudenko}.\\

 First we study the interaction of a single laser pulse with PDMS. Our numerical modeling shows that increasing the laser energy results in elongated single void formation and not a void array. Then we model the first damage structure (void) as pre-recorded into the medium and study the interaction of a subsequent laser pulse with pre-existing void. We show the generation of a second void as a result of field enhancement in front of the first void. We then model the two generated voids as pre-recorded in the medium and show the generation of the third void. Our method shows that this can continue, and successive voids can be generated as a result of multi-shot laser interaction in PDMS. \\
 
\subsection{Numerical Model and Simulation Setup}
 In our 3D numerical simulation model, Maxwell's equations are solved using the FDTD method \cite{Taflove2005} via the Yee algorithm \cite{Yee,Popov} with constitutive  relations (cgs units), $
  \textbf{H}=\textbf{B},
~
  \textbf{D}=(1+4\pi(\xi_l+\xi_kE^2)\textbf{E}$
and  current density
  $\textbf{J}=\textbf{J}_p+\textbf{J}_{PA}$. $\textbf{E}$ and $\textbf{B}$ are the electromagnetic fields, $\textbf{D}$ the displacement vector, $\textbf{H}$, the magnetic field auxiliary vector, $\xi_l$ is the linear susceptibility of the material, and $\xi_k$ is the Kerr susceptibility, which we take to be constant. The electromagnetic response of the generated plasma is represented by $\textbf{J}_p$ and  laser depletion due to photo ionization (PI) by  $\textbf{J}_{PA}$. 
 The evolution of the free electron density, $n$,  is described by: $ \frac{d n}{d t}=W_{PI}(|\varepsilon|)$, where $W_{PI}$ is the PI rate. Following Keldysh's formulation \cite{Keldysh} for the PI rate $W_{PI}$, the adiabaticity parameter for solids is $\gamma=\omega_0\sqrt{m_{eff}U_i}/eE$, where $m_{eff}=0.635m_e$ denotes the reduced mass of the electron and the hole, $U_i$ is band gap energy, $E$ is the laser electric field, and $\omega_0$ is the laser frequency.
 While we have implemented a model of avalanche ionization that follows Ref. \cite{Rethfeld},
we find that our simulation results for $70~ fs$ pulses including avalanche ionization were identical
to equivalent simulations that did not include avalanche ionization. Thus, to save computational
resources, we did not enable avalanche ionization in our code for the simulations presented here.
We assume a laser beam focused by a perfectly reflecting parabolic mirror characterized by
a given $f \#$, corresponding to laser beam waist size of $w_0 = 1.46~ \mu m$ at the focus in free space. Our model thereby eliminates the effects of spherical aberrations. The laser beam
incident onto the mirror is a Gaussian beam whose waist is half the size of the mirror.
To describe the fields focused by the parabolic mirror, the Stratton-Chu integrals \cite{Popov2009,Stratton} are
used, which specify the exact electromagnetic field emitted by the given parabolic surface. This
field is calculated on five boundaries of the 3D FDTD simulation in a total field/scattered field
framework. The laser pulses are Gaussian in
time with a pulse duration of $70 ~ fs$ and a wavelength of $\lambda = 800~ nm $ and they are linearly
polarized along the y direction and propagating along the x direction. The geometrical laser
focus is located at $x = 20 \mu m$, and the simulation domain is $50~\mu m \times 20~ \mu m \times 20~ \mu m$, with
grid size $\Delta x = \Delta y = \Delta z = 0.025 ~\mu m$. To ensure the domain was large enough, we chose it
such that, in all simulations, the laser pulse was no longer creating plasma well before it exited
the domain. Further, we ran simulations where we placed the geometric focus much deeper
within the simulation domain and found no difference in our results. 

 

The linear refractive index of PDMS is $1.5$.
The saturation density is estimated for PDMS as $n_s \approx 5 n_{cr}$, where $n_{cr} \approx 1.75 \times 10^{21}~ cm^{−3}$ is the critical electron density for the free-space laser wavelength  $0.8 ~ \mu m$. The saturation density $n_s$ is the maximum density that can be reached when every molecule is ionized. 
The FDTD  simulates linear and nonlinear laser propagation in the medium, plasma generation, ionizational and collisional energy losses as well as laser interaction with the created plasma. The accuracy of the simulations is limited by the uncertainties in some of the parameters involved. The third- order susceptibility, $\chi^3$ of pure PMMA is relatively well known and is estimated to be $3 \times 10^{-14} ~esu$ \cite{Damore2004}. The polymer used in the experiment is PDMS. $\chi^3$ of PDMS is not well known. Here we used the $\chi^3$ of PMMA; however, we show that the physical process is linear, therefore not really dependent on nonlinear susceptibility of the material. We present the results of our simulations with a $\chi^3$ that is twice that of PMMA and with Kerr effect turned off and show that the interaction is linear, therefore the overall results do not change with varying the nonlinear susceptibility of the material.\\
\begin{figure}[h]
\begin{center}
\includegraphics[width=\linewidth]{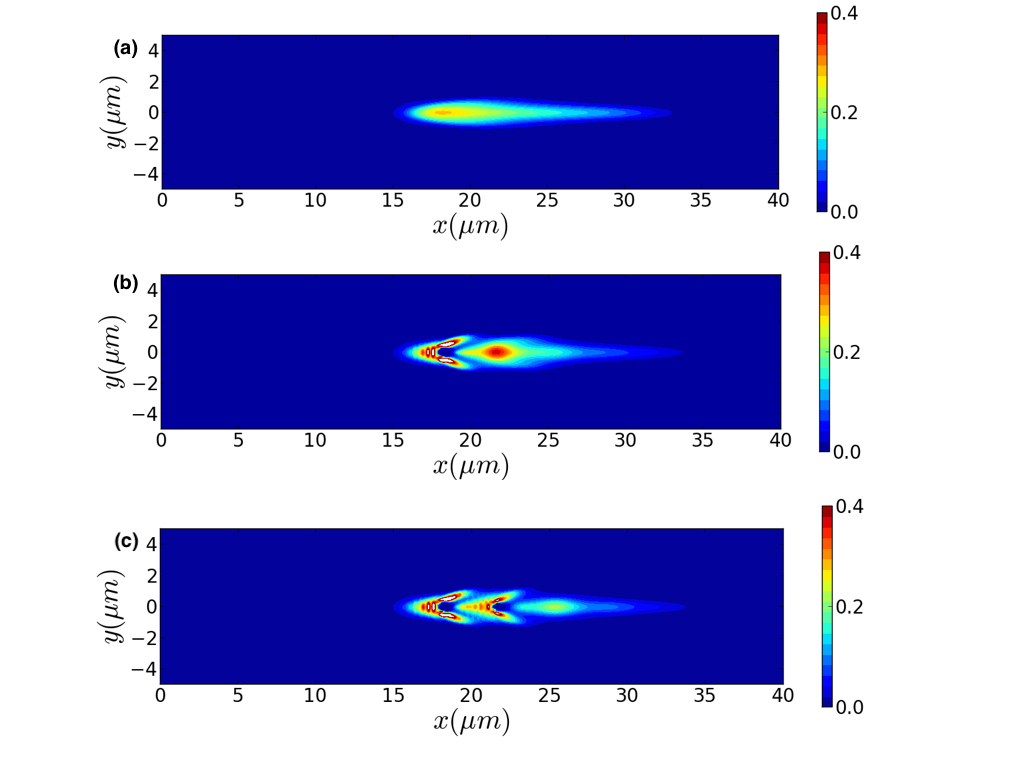}
\end{center}
\caption{ Contour plot of final electron density normalized to $n_{cr}$, (a) in bulk of PDMS, (b) with one pre-recorded void located at $x=18.4~\mu m$, and (c) with two pre-recorded voids located at $x=18.4, 22.3~\mu m$.}\label{fig1}
\end{figure}
\subsection{Simulation Results}
The mechanism of void array formation, as observed in the experiments is a multi-laser pulse effect. The first pulse generates a single void structure, and subsequent irradiation leads to the formation of multiple voids (see fig. \ref{fig1}). In order to understand the mechanism of multi-void formation, first we performed simulations of laser interaction with bulk  PDMS. The resulting electron density structure, depicted in fig. \ref{fig1}-a, closely resembles the experimental findings, featuring an oval-shaped void located at $18.4~\mu m$ with a maximum electron density of $0.25n_{cr}$. Notably, the damage threshold density for PDMS is lower than that of fused silica, and we set the cut-off damage threshold density for PDMS at $0.11n_{cr}$.The simulation also indicates that the laser focus position in PDMS is near the geometrical focus, suggesting that Kerr nonlinearity is not the primary interaction mechanism. The plasma is confined to near the geometric focus, and is rapidly formed. Plasma defocusing is also seen, but since geometric defocusing is so strong (after the geometrical
focus), it dominates over Kerr self-focusing. \\
To investigate the role of Kerr self-focusing, we conducted simulations similar to the previous simulation (fig. \ref{fig1}-a), but with the nonlinear Kerr effect was turned off by setting the Kerr susceptibility to zero. The results, shown in figs. \ref{fig2} and \ref{fig3}-a, confirm that Kerr nonlinearity is not the dominant mechanism, and that geometric focusing plays a more significant role. Figure \ref{fig2} displays the plasma density along the laser axis after the laser pulse exits the medium, with the solid and dash-dotted curves representing the results obtained with and without the Kerr effect, respectively. Figure \ref{fig3}-a presents the corresponding contour plot.

To further confirm that the interaction is primarily linear, we conducted additional simulations equivalent to fig. \ref{fig1}-a, except that we doubled the Kerr nonlinear susceptibility ($2\chi^3$). The resulting plasma density along the laser axis is displayed in fig. \ref{fig2}. The solid, dashed, and dotted-dashed lines represent the simulations with Kerr nonlinearity included, turned off, and with doubled nonlinear susceptibility, respectively. As seen in fig. \ref{fig2}, the location of the void is not significantly affected in any of the three simulations, and the plasma density is well above the damage threshold. The maximum plasma density values, $n=0.28,0.22,0.37~n_{cr}$, obtained with and without Kerr effect, as well as with twice the nonlinear coefficient $2\chi^3$, are all close to each other and above the permanent damage threshold. We therefore conclude that the Kerr nonlinear effect does not play a dominant role, and that geometric focusing is the primary mechanism. It is worth noting that increasing the intensity of the laser pulse results in an elongated void, confirming the experimental findings.
\begin{figure}[h]
\begin{center}
\includegraphics[ width=0.5\textwidth]{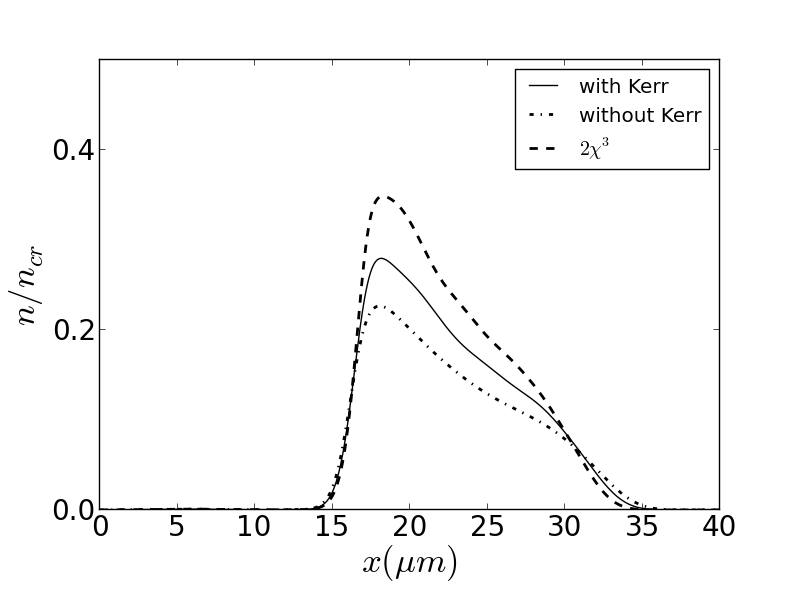}
\end{center}
\caption{On axis lineouts of final electron density in bulk of PDMS (solid curve), with Kerr effect turned off (dashed-dotted curve) and with a doubled nonlinear susceptibility $\chi^3$ (dashed curve).}\label{fig2}
\end{figure}
We have developed a novel model to better understand the mechanism behind the formation of multiple voids observed in the experiments. For a subsequent simulation representing a second laser shot, we implemented the first void (he damage structure produced by the first pulse) in the medium. This consists of a concentric sphere with an inner radius of $0.4~\mu$m and an outer radius of $0.5~\mu$m. The center of the sphere was filled with air to simulate the refractive index of the void, while the shell had a slightly higher linear refractive index than PDMS to model a densified shell ($\Delta n=0.1$). The location of the modeled void was determined based on the location where the maximum plasma density was observed in the previous simulation with the Kerr effect included $(x=18.4~\mu$m) (fig. \ref{fig1}). All other parameters were kept the same as in the previous simulation.

\begin{figure}[h]
\begin{center}
\includegraphics[width=\linewidth]{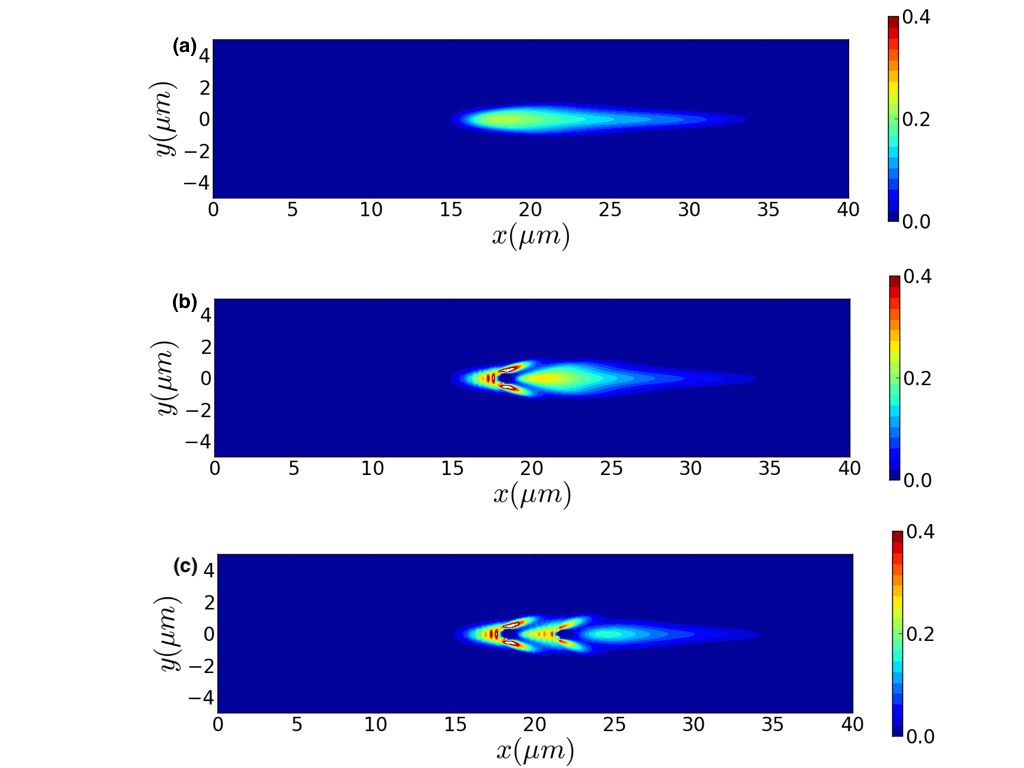}
\end{center}
\caption{Contour plot of final electron density normalized to $n_{cr}$, (a) with Kerr effect turned off in bulk of PDMS, (b) with one pre-recorded void located at $x=18.4~\mu m$, and (c) with two pre-recorded voids located at $x=18.4, 22.3~\mu m$. }\label{fig3}
\end{figure}
The subsequent laser pulse, which could be the second or third pulse in the experiment, enters the medium from the left boundary and interacts with the medium containing the pre-recorded void. We see that electric field enhancement occurs to the right of the pre-recorded void, leading to the formation of a second void. The plasma density contour plot after the laser pulse has left the medium is shown in fig. \ref{fig1}-b. A portion of the laser pulse scatters from the void and focuses in front of it, at a distance of $3.5~\mu$m from the center of the void, where the laser intensity is sufficient to ionize the PDMS medium, resulting in the formation of a second void at a distance of $21.8~\mu$m with a radius of $0.4~\mu$m. Furthermore, we find plasma density above threshold at the perimeter of the pre-existing void, suggesting that void sizes grow with increasing number of shots, as we observed experimentally  (see figs. \ref{fig2e},\ref{fig3e}). To verify that this mechanism is linear, we performed simulations similar to those in fig. \ref{fig1}-b but with the nonlinear Kerr effect turned off by setting the Kerr susceptibility to zero. The resulting plasma densities along the laser axis are shown in fig. \ref{fig4}. The position of the laser focus in front of the void is almost identical to that in the simulation with the nonlinear Kerr effect included, and the maximum electron density is also similar. Therefore, we conclude that the interaction of the second (or third) laser pulse with PDMS is a linear effect, and the leading mechanism is geometrical focusing.\\

\begin{figure}
\begin{center}
\includegraphics[ width=0.5\textwidth]{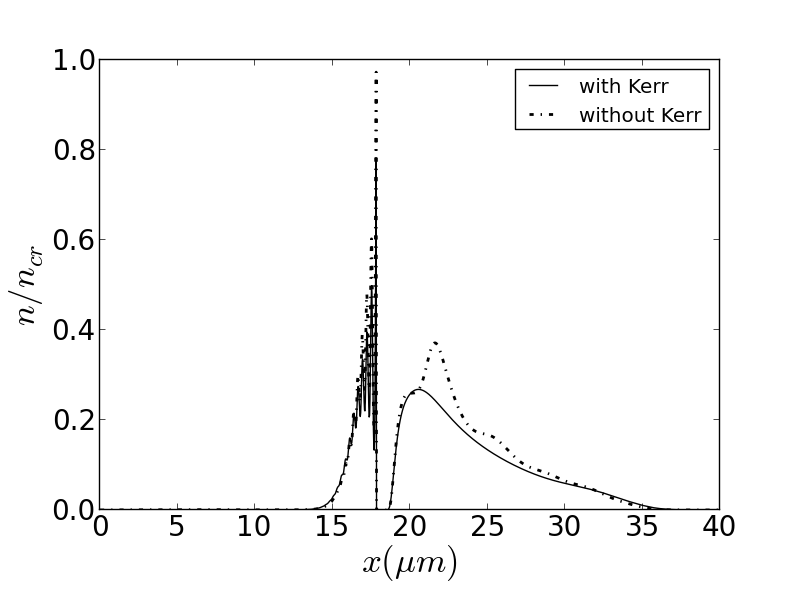}
\end{center}
\caption{ Lineouts of on axis electron density normalized to $n_{cr}$ in the middle $x-y$ plane. Solid and dashed curves correspond to simulations without Kerr nonlinearity and with nonlinear Kerr effect, respectively. A spherical void is located at $x=18.4~\mu m$.}\label{fig4}
\end{figure}
To verify that the void generation continued with subsequent laser pulses, we modeled both the first and the second voids as concentric spheres with refractive index of air in the centers, and densified shells with slightly higher linear refractive index than the medium ($\Delta n=0.1$). The modeled voids were placed at the locations where the first $(x=18.4~\mu m)$ and the second void $(x=22.3~\mu m)$ were observed (fig.\ref{fig1}). Figure \ref{fig1}-c illustrates the interaction of the laser with two voids previously recorded.  The contour plot of the electron density shows the generation of the third void in front of the second void, where field enhancement occurred in the right side of the second void. The radius of the third void was $0.4~\mu m$, and the electron density was above the damage threshold. The process of void array generation in PDMS can continue by adding the voids one after another using the method presented here.\\
The mechanism presented in this study clearly illustrates the successive generation of voids in PDMS as a result of multi-laser pulse interaction. To confirm that nonlinearity is not the leading process in void generation, we conducted simulations with one and two voids while turning off the Kerr nonlinearity and keeping other parameters the same as in fig. \ref{fig1}-c. The electron density contour plots with nonlinear Kerr effect turned off are shown in figs. \ref{fig3}-b,c. As observed before, the second and third voids are generated in front of the first and second void, respectively. The on-axis lineouts of electron densities in fig. \ref{fig5} (solid curves) indicate that the second and third voids are located very close to the simulation result including Kerr effect (dashed curves). Additionally, the maximum electron densities at the position of the second and third voids are still above the damage threshold. Therefore, the mechanism is primarily linear. 
\begin{figure}
\begin{center}
\includegraphics[ width=0.5\textwidth]{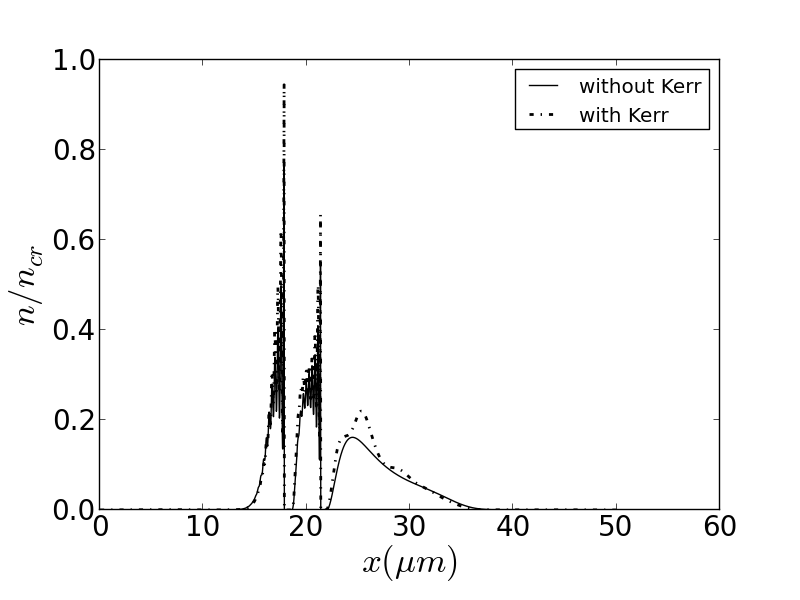}
\end{center}
\caption{ Lineouts of on axis plasma density normalized to $n_{cr}$ in the middle $x-y$ plane. Solid and dashed curves corresponds to simulations without Kerr nonlinear  and with nonlinear Kerr effect, respectively. Two spherical voids are located at $x=18.4, 22.3 ~\mu m$ and $x=22.3~\mu m$.}\label{fig5}
\end{figure}
\section{Conclusion}
In conclusion, we studied the mechanism of laser induced void array formation in PDMS.
Our experimental results shows that laser energy, number of pulses per micron (writing speed) and laser spot size (NA) affect the multi-void formation in PDMS: multi-void formation happens as a result of multi-pulse irradiation in bulk of PDMS. In addition, our results demonstrated that increasing laser energy and number of pulses per micron for fixed laser parameter results in increasing the number of voids as shown in fig. \ref{fig2e}.  We also showed that tighter focusing conditions (higher NA) leads  to smaller size of voids, with shorter distance between the voids.  The simulations presented in the study have reproduced the generation of void arrays in PDMS using a similar multi-laser pulse approach. By modeling the voids  produced by previous laser pulses as concentric spheres with densified shells and simulating the laser interaction with these pre-existing voids, we showed that additional voids are created.\\
To understand the process of multi-void formation in PDMS, we performed high-resolution Finite-Difference-Time-Domain (FDTD) simulations. Initially, we focused on the interaction of laser with the PDMS bulk. Using the same parameters as those used in the experimental setup led to the formation of a single void structure close to geometrical focus of the laser pulse.
Subsequently, we implemented a pre-recorded void as concentric sphere with densified shells and simulating the laser interaction with the void, we demonstrated the generation of a secondary void as a result of laser interaction with pre-recorded void. To further validate this mechanism, we conducted additional simulations using two pre-recorded voids, which were implemented as concentric spheres with densified shells in the medium. As expected, when the laser interacted with these voids, it led to the formation of a third void in front of the second pre-recorded void.To explore the underlying mechanism further, we conducted simulations with the Kerr nonlinearity terms turned off in simulations. A comparative analysis of the simulations, with and without the Kerr effect showed that the leading process governing the multi-void formation in PDMS is a linear mechanism.\\

\section*{Acknowledgements}
This work was supported by the Canadian Foundation for Innovation, the Digital Research Alliance of Canada (formerly Compute Canada), the Canada Research Chairs Program, and Air Force Office of Scientific Research (FA9550-14-1-0247). 
A. M. Alshehri would like to thank King Khalid University for funding this
work through the Research Group Program under grant
number RGP 1/241/44.\\
\section*{Author contributions statement}
A. M. Alshehri and R. Bhardwaj conducted experiments and analyzed experimental results.
N. Naseri performed simulations, generated figures and prepared manuscript. N. Naseri and L. Ramunno analyzed and discussed the simulation results.
N. Naseri, L. Ramunno, A. M. Alshehri and R. Bhardwaj discussed and compared the experimental results with numerical results. All authors reviewed the manuscript. \\

\section*{Additional information}

The datasets used and analyzed during the current study available from the corresponding author on reasonable request.

\end{document}